# Analysis of the Actual Scientific Inquiries of Physicists I

## - Focused on research motivation -

**Park, Jongwon***
**Department of Physics Education, Chonnam National University,
Gwangju, 500-757**
**Jang, Kyoung-ae****
**Center for Educational Research, Seoul National University, Seoul, 151-742**

**ABSTRACT**
This study was investigated to understand the in-depth features and processes of physicists' scientific inquiries. At first, research motives were investigated by interviewing six physicists who were prominent worldwide. As a result, three main types - incompleteness, discovery, and conflict - and nine subtypes of research motivation, were identified. Six additional background factors were found which might affect the design and start of research. Based on these findings, implications for teaching scientific inquiries to students were discussed.

**I. Study Background**

Science learning through scientific inquiries can help students construct and develop scientific knowledge, improve scientific process skills needed for scientific investigations, and encourage students' motivation for learning science [1].

Therefore, one of the main issues in teaching scientific inquiries is to understand the authentic process of how scientific inquiries are structured. Park (2004) recently suggested the model of scientific inquiries (Fig. 1) and discussed how students acted in each stage of the inquiries [2]. Studies in each stage included the analysis of students' behaviors when observing scientific phenomenon [3], generating scientific hypotheses [4], designing experiments to test hypotheses [5], explaining natural phenomenon [6], and evaluating scientific evidence [7]. In this model of scientific inquiries, Park (2004) assumed that scientists perform the following activities [2]:

- to discover specific features or suggest generalized rules through scientific observation about natural or laboratory phenomenon.
- to generate scientific hypotheses to explain new phenomenon and to test the hypotheses experimentally.
- based on scientific theories and laws, explain or predict natural phenomenon or other theories and laws.
- to re-interpret or organize the established scientific theories or laws, and to apply them to various situations for inventing new products or solving complex problems.

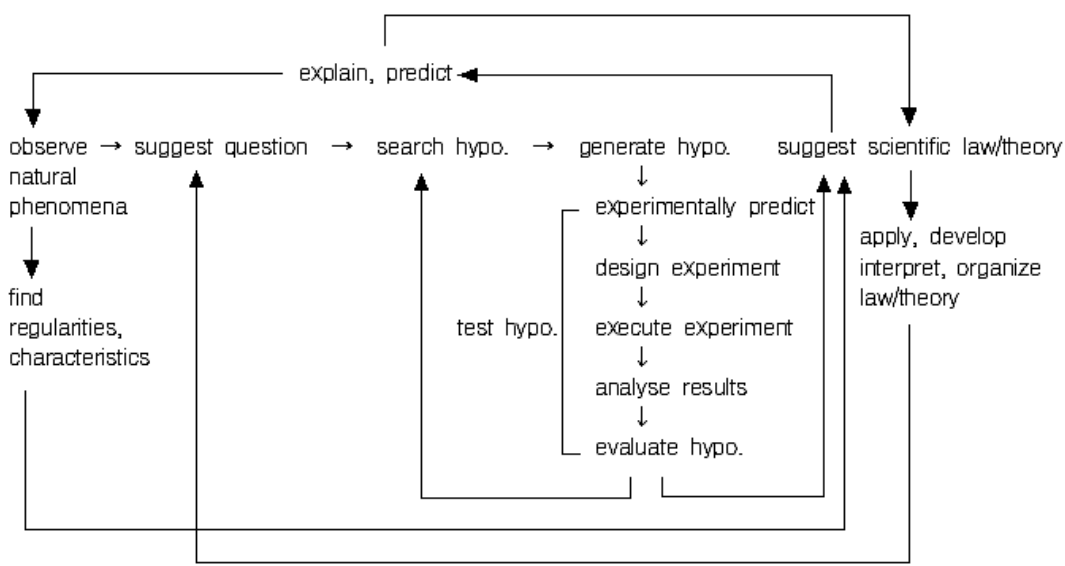

<Fig. 1> Process of Scientific inquiries

Then, do scientists actually perform research according to the processes as in Fig. 1? This study was initiated from this question.

Recently, many researchers criticized the reasoning processes involved in school science textbooks as qualitatively different from the processes employed in real life scientific inquiries [8]. For instance, Chinn and Malhotra [9] point out that, contrary to the oversimplified forms of inquiries used in schools, actual scientists select their own variables or often invent variables before conducting experiments. They construct procedures using an analog model, measure different variables, spend time checking out possible errors in procedures, seek out global consistency within a complex web of data and theories, and so on. Hmelso-Silver et al. [10] also summarized the major features of activities relating to inquiries of scientists. For instance, scientists spend a great deal of time constructing and identifying a research problem before starting an experiment and devote their effort to more theory-driven experimental designs than data-driven ones.

Scientific inquiries also take place in theoretical research. For instance, in actual scientific situations, theoretical research goals are meant to develop and refine theoretical models in response to evidence [11, 12]. This helps pursue the logical consistency within the theory [13, 14], or gives a logical explanation to other theories as well as to a natural phenomenon [6, 15].

Likewise, the previous studies imply that our understanding about scientific inquiries in schools may differ from scientific inquiries by actual scientists. However, studies to identify and understand the features and methods of scientific inquiries by scientists are not yet sufficient.

Therefore, we need to have a detailed and systematic understanding of the cognitive processes used by scientists in scientific research. By understanding this, we can obtain implications/inferences for teaching scientific inquiries to students in more authentic ways. Our research was initiated from the following questions:

(1) What research motives do scientists have in mind when they start scientific inquiries?

(2) What process skills or research strategies do scientists employ when they engage in scientific inquiries?

(3) Which types and what features are there in research results obtained by scientists' scientific inquiries?

Before starting this study, two points are worth mentioning. The first one is that even though the main interest of this study is to identify and to understand the features and processes of scientific inquiries of scientists, our study did not follow the classical view of 'pupil as scientist'. The classical view of 'pupil as scientist' assumes that learners can construct or discover scientific knowledge when they engage in a scientific inquiry just as the procedures of actual scientists do. The basic goal of this view is to make students act like scientists for a day [16]. However, it has been well known that classical view of 'pupil as scientist' has various limitations and has epistemological distorted basis. For instance, this view assumes that if students carefully explore natural phenomenon for themselves, then they can reach scientific understanding by organizing facts obtained by exploring nature. But, many researchers have pointed that this assumption is not correct in the context of learning science in schools [17-19]. Here, an important point is to make students "scientifically literate" through authentic scientific inquiry. The authentic scientific inquiries for students can be constructed by having a greater understanding of the characteristics and procedures of scientific inquiries that scientists conduct [20 -22].

Another point worth noting is that some researchers have stressed there does not seem to be a consensus on the process of scientific inquiry. That is, different scientists in different areas employ different methods and procedures when conducting a scientific inquiry. Therefore, scientific inquiries do not follow a definite set of rules or algorithmic procedures requiring particular actions at particular stages of the scientific inquiry [8, 23]. This view is partially supported by this paper. However, the basic goal of investigating main features and processes of scientific inquiries by scientists, is not to construct algorithmic and fixed models of scientific inquiries, but to create a starting point for teaching scientific inquiries to students.

**II. Purpose of This Paper**

Our main concerns about scientific inquiries are: (i) the motivation of scientist's research; (ii) the processes of scientific inquiries and process skills they use; and (iii) types of results obtained by a scientist's research.

As a starting point of our project, this paper will present only the results about the first research concern; that is, types and features of research motivation of physicists. Other results about the second and the third research questions will be presented in successive papers.

Research motivation for physicists has special importance to various aspects of scientific inquiry to students. Since motivation in cognitive aspects rather than affective ones are closely linked to research purposes, understanding about the cognitive research motivation can lead us more directly to encourage students to conduct scientific inquiries by themselves.

Regarding the research motivation of physicists, Zichichi (1999) [24] summarized five factors which can lead to original work in science: (i) technological inventions in experiments; (ii) suggestion of new ideas in physics; (iii) new projects for large physics laboratories; (iv) international collaboration for new scientific culture; and (v) problems related to Planetary Emergencies (e.g., problem of clean energy). Here, we see that the environmental factors act as a motivation for research.

Wilson[25] briefly described what makes scientists choose certain research problems. Namely, scientists choose specific research problems when they have strong interests in a field, new ideas or theoretical new tools, professional experiences or unique equipment from colleagues. Research topics also can be chosen when there is a untouched or uninvestigated region, or when the topics have special significance. Wilson's brief descriptions give interesting implications for research motivation, but he did not present concrete evidence related to those suggestions.

Except for the aforementioned literature, it is not easy to find out a scientist's research motives. Moreover, there is scant research using interviews with scientists. Also, no research has been discussed about the students' scientific inquiries based on implications drawn from the results about physicist's research motivation.

In this paper, through interviews with physicists, the features of research motivation will be identified. As supporting evidence, passages from papers by interviewees and instances from science history will also be described. Finally, implications for teaching scientific inquiry to students will be discussed.

**III. Procedures**

**Participants**

In this paper, three theoretical physicists and three experimental physicists participated. They are all well-known academically, and are presently conducting research in particle physics, nuclei physics, or condensed matter and solid-state physics.

**Interview**

Interviews were conducted with each physicist for about 1-1.5 hours. Questions were asked in the interview about research motivation, procedures or inquiry skills used during research, and the main results obtained by the research (see Table 1). Research topics were selected by participants who had outstanding papers or research projects which were completed or were still under investigation. All interviews were audio recoded and analyzed. In this paper, only the responses to the first question in Table 1 will be presented.

Table 1. Major Interview Questions
-------------------------------------------------------------------------
1. Motivation of research
    (Where was your research initiated?
2. Research procedures or methods
    (Can you explain the processes of your research?
3. Main results of research

(What were the main results of your research?
\------------------------------------------------------------------------

After the interview, participants recommended papers related to the topics of interview, then we reviewed and analyzed the papers to obtain additional information related to the major questions.

## IV. Results from the first Research Question

### 1. Types and Features of Research Motivation

When we asked about the main research motivation, their responses were classified into 3 main types and 9 subtypes (Table 2).

**Table 2. Types of Research Motivation**
\------------------------------------------------------------------------

M1: Incompleteness
    M11: Inaccuracy of experiments
    M12: Unidentified/undeveloped areas
    M13: Unproven parts of the theory
M2: Discovery & Development
    M21: Discovery of new data/phenomenon
    M22: Suggestion of new theory
    M23: Development of new materials
    M24: Development of new experimental techniques/equipments
M3: Conflict
    M31: Mismatches between theory and experiment (Unexplainable phenomenon)
    M32: Internal conflicts inside the theory
\------------------------------------------------------------------------

**The First type of Research Motivation: Incompleteness**

According to interviews and reviews of academic papers, we found that the first type of research motivation was 'incompleteness'. When accurate physical quantities were not obtained because of noises while obtaining data or incomplete measuring techniques, physicists wanted to start researching to obtain more accurate and complete data (M11).

Unidentified or undeveloped area can also play a role in research motivation (M12). This is the case when physical quantities or structures for specific new materials are not yet measured, certain physical phenomenon in specific conditions (e.g. on a smaller scale or higher temperature) are not analyzed yet, or some field of physics is not investigated or developed yet. Regarding these motives, Shamon [26] described as follows; *"Following J.J. Thomson's identification of the electron in 1897, it was natural that attempts would be made to determine precisely its properties. Thomson had measured the ratio of charge to mass of this elementary particle, thereby showing the ratio, at least, to be unique. The next obvious step was the determination of its mass or charge separately."* (p. 238)

Especially, in the area of theoretical research, it was found that unproved assumptions or incomplete theories could also function as motivations for research (M13). For instance, Einstein said, *"I was dissatisfied with the special theory of relativity, since the theory was restricted to frames of reference moving with constant velocity relative to each other and could not be applied to the general motion of a reference frame. I struggled to remove this restriction and wanted to formulate the problem in the general case."* (pp.244-245) [27]. Relating to the first type of research motivation (M1), the followings are the passages transcribed from the papers or responses from the interviews.

*"The main disadvantage of ScI crystal would be its high internal background from cesium radioisotopes. However, there has been no accurate measurement of the internal background level and no serious attempt at internal background reduction for CsI crystal." (M11: K-paper)*

*"I want to solve the problems that physicists have not solved yet. ... Physics applied to one dimension may differ (from physics applied to other dimensions) ... in the area nearby 50 or 100 nano-scale, some strange events happen." (M12: G-interview)*

*"(This) has not been investigated by any physicists. So we submitted a research proposal." (M12: P-interview)*

*"... we have used very complex theory called quantum chromodynamics. This (theory) can be used clearly and simply in the very high energy level, but we can not solve the problems in the low energy like everyday conditions..." (M12: M-interview)*

*"... naive factorization was assumed, ... But there was no justification for this assumption except the argument of color transparency." (M13: C-paper)*

**The Second type of Research Motivation: Discovery & Development**

New scientific research can be initiated from previous scientific discoveries and development. For instance, when new physical phenomenon are observed, new materials developed, or new material structures uncovered, physicists can start new investigations to explore these findings (M21). For instance, Gribbin [28] described how the discovery of a new phenomenon led to new research; *"...he(Becguerel) head of the hot news about X-rays, including the discovery that they originated from a bright spot on the glass wall of cathode ray tubes, where the cathode rays struck the glass and made it fluorescence. This suggested to him that phosphorescent objects, which also glow in the dark, might produce X-rays, and he immediately set out this hypothesis ..."* (p.496).

When a new theory is suggested, physicists usually focus on the experimental evidence that confirm the new theory. And physicists also derive new unobserved experimental predications or clues for solving other problems from the new theory. Therefore, a new theory would act as a starting point for new research (M22).

Einstein also mentioned that Lorentz's solution was the motive of studying Fizeau's experiment; "*I had a chance to read Lorentz's monograph of 1895. He discussed and solved completely the problem of electrodynamics with the first [order of] approximation ... Then I tried to discuss the Fizeau experiment on the assumption that ...*" [27]

Likewise, the development of new materials or experimental techniques can lead to new research motives (M23 and M24). Gribbin [28] also stressed the importance of experimental developments in physics as follows: "*The biggest revolution in the history of science began with the invention of a better kind of vacuum pump, in the middle of the nineteenth century.*" (p. 487). Wilson [25] mentioned that, "*the field of microwave spectroscopy has always been an attractive one, but until the invention of magnetron and klystron oscillators, it could not be exploited.*" (p. 1) Following are examples included in the second type of motives for research (M2).

"*Recently, it has been shown experimentally that fullerences or endohedral metallofullerences ... can be inserted into single-wall nanotubes (SWNTs), forming a pea pod-like structure." (M21: G-Paper)*

"*Two or three years ago, it was published that carbon nanotubes could be used to make an electric element such as switch theoretically. So I started this research to make it experimentally." (M22: P-interview)*

*"My graduate student said that there was a good paper (which could be used to solve our problem)... we tried to test whether it would generate a bad point like our (previous) case ... but it did not generate (a bad point) ... After all, we found coherent theory." (M22: C-interview)*

*"When new materials appeared (discovered), we conducted experiments to identify its properties, ... and if new experimental tool (equipments) were developed, by using these new equipments, we could find out something new which were not known previously." (M23, M24: P-interview)*

**The Third Type of Research Motivation: Conflict**

If a physicist finds out that there are any inconsistencies between theory and experimental data (M31), or there are any internal incoherencies or conflicts inside the theory (M32), these cases become highly important for motivating research. For instance, Plank's new research was initiated by this discrepancy; "*By the summer of 1990, ... discrepancies had been discovered between data and formula that could not easily be explained away as errors of measurement. ... (Rubense called on the Planks) Rubens described to Plank new findings that clearly violated Wien's law. ... Before he (Plank) went to bed that night, Plank had conjectured a new formula to replace Wien's law, ...*" (p. 49) [29]

Controversy or conflict between two theories can also act as a motive for new research. For instance, G.P Thomson, the son of J.J. Thomson, described the background of the discovery of the electron as follows: "*J.J. Thomson turned to the line which led to the discovery of the electron, namely cathode rays. Cathode rays had been known for about 50 years, but there was at the time, a great controversy as to their nature, ... waves, ... particles.*" (p. 291) [30]

Many other examples exist showing that anomalies or conflicts cause paradigm shifts in physics [31]. Following are extracts from Kuhn's writing, the papers by interviewee, and responses to the interview.

*"Discovery commences with the awareness of anomaly ( i.e, with the recognition that nature or experimental data/phenomenon, has somehow violated the paradigm-induces expectations- theoretical prediction), that governs normal science. ...In both cases the perception of anomaly --- played an essential role in preparing the way (starting new research) for perception of novelty." (pp. 52-53, 57)*

*"Especially the treatment of non-perturbative effects from the strong interaction is a serious theoretical problem in non-leptonic decays. Precise experimental observation of non-leptonic B decays make it urgent to give firm theoretical predictions including the effects of CP violation." (M31: C-paper)*

*"... when we are unsatisfied with the previous theory. If we found that there are inconsistencies inside the theory, we have to check out what we misunderstand ... or if there is no problem in our understanding, we have to make a new theory." (M32: C-interview)*

**Implications for Teaching Scientific inquiries**

Recently, 'open' scientific inquiries have been stressed as an better approach for teaching scientific inquiry to students [32]. In the case of open scientific inquiries, students are encouraged to set up inquiry problems themselves with internal motive. Unlike affective motives such as interest or curiosity in a topic, our findings demonstrate cognitive and intellectual aspects. Additionally, three types of research motives are closely related to research purposes. Therefore, helping students to have cognitive motivation may directly guide students to have concrete inquiry problems. However, teaching or guiding research motivation for students has not been treated appropriately in ordinary schools. Therefore, we need to be concerned about how to help students be cognitively motivated for designing scientific inquiries.

Regarding incompleteness, the first type of research motivation, many students think that scientific knowledge, concepts, or laws are complete. That is, they regard scientific knowledge as absolute which no longer to be modified, refined, or developed [33]. If students recognize scientific knowledge has built-in limitations, unproven assumptions, and inaccuracies, then this reorganization can play a role in motivating scientific inquiries. For instance, when students investigate Ohm's Law, they may feel something is strange when a resistor is heated or when it is cooled down by putting it inside liquid nitrogen. That is, they can recognize that they have no understanding what happened in those situations. Then this perplexity might act as a motive for future investigation by the students.

Likewise, the recent introduction of discoveries and developments in physics during scientific inquiries helps students with a motive for exploring further. For instance, when carbon nanotubes were discovered in 1991, their value and utilization have been expanded to various fields. Thus, by introducing extensive applications for carbon nanotubes such as their use in semiconductors or probing atomic microscopes, students can be encouraged to realize how discoveries can lead to various inventions.

Regarding the third type of research motivation (reorganization of conflicts between theory and experimental results), has been/is emphasized as a crucial condition for changing student's prior misconceptions [34, 7, 6]. This is the same in scientific inquiries as conceptual changes. Therefore, introducing experimental evidence, phenomenon, or results which can not be explained by students, or helping a student recognize logical inconsistencies in their own reasoning, acts as an important role in intellectual motivation for students to solve conflicts through scientific inquiries.

As a result, three types of research motives by physicists need to be included in the process of a student's scientific inquiries.

**2. Background Factors Affecting Motivation of Research**

From the interviews, we obtained additional factors, which could impact he start of research. These factors are not cognitive, but affective, attitudinal, or environmental aspects (Table 3).

**Table 3. General Backgrounds Factors Affecting the Start of Research**

---

B1: Attitude as a problem generator rather than problem solver

B2: Read many recent papers and obtain relevant information

B3: Interact with various other research fields

B4: Recognition of future prospects of research

B5: Recognition of scientist's own talents and abilities

B6: Various (financial) support

---

At first, interviewees mentioned that in order to be a creative researcher, it was important to adopt an attitude as a problem generator, rather than problem solver (B1). Many scientists have emphasized this. For instance, Einstein said, "*the formulation of a problem is often more essential than its solution*"[35]. In fact, problem finding has been recognized as an important indicator in the theory of creativity [36].

To generate and set up new problems that are valuable for research, interviewees noted that reading relevant papers was essential (B2) to obtain valuable and important information. Additionally, obtaining new ideas by interacting with other research fields was also important (B3).

Factor B2 means that a deep understating of previous scientific knowledge is important for inventing new knowledge [4, 37-39].

Factor B3 is also noted as one of the important strategies for creativity. For instance, Park [2] suggested 'the associational thinking' as a component of creative thinking. This thinking can be defined as reasoning connecting new information to existing knowledge, combining two ideas, which seem to be separate with each other, or uniting two opposing concepts. Similarly, thinking of conceptual combination has also been stressed an important factor for generating problems as well as solving the problem [40,41]

*"Through discussions with students in doctoral courses, I tried to generate new problems again and again which could require about 4-5 years to solve it..." (B1: M-interview)*

*"(Let's imagine) there is a person who tries to make a wheel as an invention in a temple (without any interaction with outside world). He does not know that a wheel already exists in the world. He should have known that fact. ... That is, we have to know clearly what have done up to now. .. To do this, I usually read many recent papers everyday." (B2: G-interview)*

*"When we treat a one-dimensional problem, ... there are two stories (which can be combined)... one is about the virus ... in some situations, if the center of the virus is empty, then it (virus) can easily accept the metal ... then a one dimensional metal can be generated inside the virus ... this idea comes to mind by talking with biologists..." (B3: G-interview)*

*"... by contacting others who are conducting research in other fields or who have other (research) backgrounds, we search (for) possibilities which can generate synergy effect." (B3: H-interview)*

Besides the recognition of recent developments in physics research, some interviewees mentioned that recognizing future prospects in physics research will be an important factor for planning (B4). On a related point, Jang [42] found that many scientists mentioned social prospects (such as development possibilities), as one of the most influential factors when selecting a job.

Another interesting factor affecting research plans was the recognition of a scientist's abilities in a specific area (B5). That is, physicists usually considers which fields are the most competitive to them, or deliberate whether they would draw the highest level of result before starting the research.

*"We have to find out, among our concerns, which area can be the most competitive to us." (B5: K-interview)*

*"My ability is also important. ... I should have the greatest competitive power in (the) world. In some cases, I abandon starting the research even though I have interests in that research ... because I would not be one of the strong competitors in a world (in the field of that research)." (B5: G-interview)*

Finally, interviewees mentioned that various financial supports were also important for their research (B6). This means that scientific endeavors also are the social activity. As mentioned earlier, social aspects are also important factors for scientific advance. For instance, large research projects, such as, the ITER (International Thermonuclear Experimental Reactor) or the HGP (Human Genome Project), were possible by investing large amounts of capital including generous grants for research staff.

*"It (research) is closely connected to realistic problems. Even though it (research projects) (are) actually (a) very good idea, it cannot be executed without financial support." (B6: K-interview)*

**Implications for Teaching Scientific Inquiries**

The factors that affect research motivation were found to closely relate to creativity in science. Therefore, these factors needed to be encouraged when teaching scientific creativity.

We always have stressed that students should have scientific attitudes. Accordingly, we need to encourage students to be "problem-generator thinkers", acquiring information about what has been done in the field of physics. They must try and connect what they know in physics with knowledge from other fields or explore future prospects.

Of course, to do this, more concrete teaching strategies will be needed. For instance, to help students generate problems, Park's seven strategies for finding new inquiry problems can be utilized. When presenting certain experimental results, students can suggest new inquiry problems by: (i) briefly analyzing the presented data; (ii) relating findings with what they already know: (iii) replacing equipment, materials, or methods used for obtaining data with alternatives; (iv) inferring reasons or variables which can impact on data; and (v) considering what the purpose of the experiment is, or expanding measurements [2].

Regarding B3, to link two concepts that seem to be unrelated with each other [6], following activities can be used:

1. Find many similar properties between Faraday's Law and Hooks' Law

2. Relate the index of refraction and the speed of a falling object

For the first activity, students can respond with: (i) two shapes of coils and springs used in Faraday's Law and Hook's Law are respectively similar to each other; (ii) both laws say that nature try to revert to their original state; or (iii) the two laws refer to electromagnetic phenomenon because restoring force in Hook's Law is generated by electric attraction or repulsion between molecules.

For the second task, students can respond with: (i) the value in both cases is determined by characteristics of medium; or (ii) the two are related to velocity because the index of refraction is defined by the ratio of velocities in and out of the medium.

By using these questions, we let students think twice about the concepts, letting them find new properties that are not identified.

Concept Maps$^{TM}$ developed by Novak [43] also emphasizes the relationship between concepts by connecting them using linking words (Fig. 2). Using concept maps, students can organize concepts and construct a network of concepts. Concept maps enable students to recognize how concepts are related to other concepts, finding new meanings.

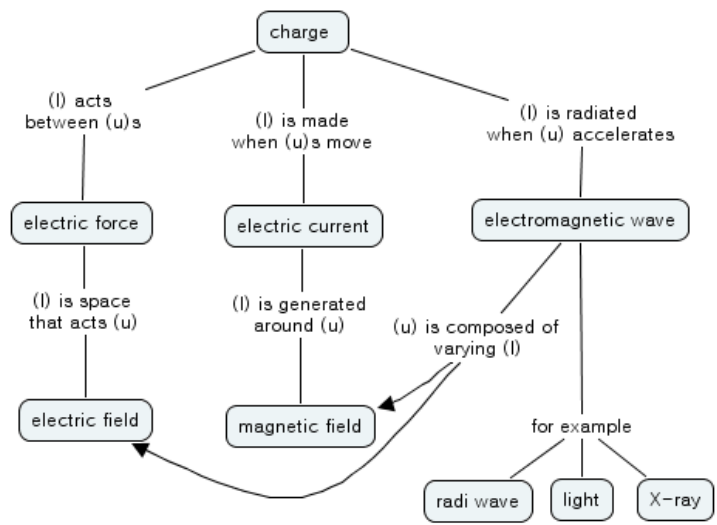

Figure 2. Concept map about electricity and magnetism.
Here, (u) indicates upper concept; (l) indicates lower concept

## V. Conclusion and Further Studies

This study was initiated to understand the nature of scientific inquiries used by actual physicists. Broken down into three aspects, these include features and types of research motivation, characteristics and processes of physicists' inquiries, and the main results of their research. This article presents findings from the first part of this study. As a result, we identified three main types of research motivation (incompleteness, discovery, and conflict), including nine subtypes. These categories were supported by instances in the history of physics, interview responses obtained from physicists, and review of research papers written by them.

Motivation usually concentrated on the affective aspect. But, the cognitive aspect of motivation also has an important role in scientific inquiries, because it may be closely related to the purpose of inquiry. Therefore, we need to encourage students to have cognitive motivations as well as affective ones. However, cognitive motivation has not been taught appropriately in the context of school laboratories.

Based on findings from this article, several recommendations for teaching research motives were suggested. Using these recommendations and suggestions, it is hoped that the results of this study can be utilized for teaching scientific inquiries to students in a more authentic way.

This article is the first part of a larger study consisting of three parts. In the next article, we will present the results about: (i) how actual physicists use scientific inquiry skills; (ii) describe the processes of physicists' scientific inquiries; (iii) explain what features there are in the inquiry process; (iv) what the main results of scientific inquiries are; and (v) how the results relate with the research motivations found in this article.